\documentstyle[12pt]{article}
\newcommand{\rms}{\rm\scriptsize}
\def\be{\begin{equation}}
\def\ee{\end{equation}}
\def\ba{\begin{eqnarray}}
\def\ea{\end{eqnarray}}

\begin{document}
%
%  Preprint Number
\noindent hep-ph/9505327
\hfill FTUV/95-21 \\ \mbox{}\hfill IFIC/95-21 \\  \mbox{}\hfill May 1995
\vspace*{1cm}
\begin{center}
{\bf\huge Constraining new interactions \\ with leptonic
tau decays}\\

\vspace{0.5cm}

{\Large Antonio Pich and Jo\~ao P.\ Silva}\\

\vspace{0.3cm}

Departament de F\'\i sica Te\`orica
and IFIC \\  Universitat de Val\`encia -- CSIC\\
Dr. Moliner 50,
E--46100 Burjassot, Val\`encia, Spain
\vspace*{1cm}  %\vspace{2cm}
\end{center}

\begin{abstract}
The recent measurements of the Michel parameters in tau decays
enable, for the first time, a
thorough analysis of the leptonic sector.
In general, in models beyond the Standard Model,
these parameters will be
altered through changes in the W and Z couplings, and/or through
interactions mediated by new gauge bosons.
We perform a complete,
model independent analysis of the constraints imposed by
the present data on such boson-mediated interactions, and point
out the existence of useful relations among the couplings.
\end{abstract}

% \newpage

\section{Introduction}
In any theory in which the fermions have interactions mediated
by heavy (scalar and/or vector) bosons, the low-energy consequences
can be conveniently parametrized by four-fermion interactions.
Hence, precise low-energy tests of processes involving fermions
constitute a window through which one may peek into the nature
of the interactions at higher energies. This procedure is,
in principle, cleaner in the lepton sector where it is not
obscured by hadronization.
Furthermore, one usually expects the scalar-mediated
interactions to have the fermionic vertices
proportional to the fermion masses, thus making the tau decays the
ideal  system for their study.

The exciting new experimental results in leptonic tau decays
reported recently \cite{ALEPH:95,ARGUS:95,ARGUS:95b}
provide most of the missing pieces of information
and warrant for the first time a complete analysis of the
lepton sector.

In the Standard Model (SM), the quantum numbers of the fermions under
SU(2)$_L$
are judiciously chosen in order to obtain a low-energy
``(V$-$A) $\otimes$ (V$-$A)" four-fermion structure,
correctly describing
the dominant features of the experiments in $\beta$ and $\mu$ decays.
One has now the opportunity to test this scheme in tau decays.
Should any difference arise, that will be a sign of
Physics Beyond the SM.
In most extensions of the SM, these new effects arise through
differences in the couplings to the $W$ and $Z$ bosons,
or through the exchange of new intermediate bosons.
The new four-fermion interactions thus obtained will be
typically dominated by a single intermediate boson;
either the one with the smallest mass
or that whose couplings to the leptons are specially large.
In any case, important relations exist between the low energy
parameters.

This program is undertaken in what follows. In section 2
we set up the analysis in terms of the helicity projection form
of the four-fermion interaction pointing out the most salient
model-independent features. We do this for completeness and to
set up the notation for the subsequent sections.
In section 3 we summarize the experimental situation and in section
4 we discuss the universality tests.
Section 5 is devoted to the analysis of non-standard charged
intermediate bosons and section 6 to lepton-flavour-changing
neutral-boson interactions.
Section 7 contains a summary of some features of our analysis and
resulting information on the opportunities for physics beyond the SM.
Finally, we draw our conclusions in section 8.
The appendix is devoted to the development of relations relevant for
the analysis of lepton-flavour-changing neutral bosons and
a detailed discussion of the consequences of the unobservability
of the final-state neutrinos.
We also discuss there the complementary information extractable from
neutrinoless charged-lepton decays.

\section{The four-fermion hamiltonian}

Let us consider the leptonic decays $l^-\to\nu_l l'^-\bar\nu_{l'}$,
where the lepton pair ($l$,$l^\prime $)
may be ($\mu$,$e$), ($\tau$,$e$), or ($\tau$,$\mu$).
The most general derivative-free, lepton-number conserving, four-lepton
interaction hamiltonian,
consistent with locality and Lorentz invariance,
can be written as \cite{scheck}
\be
{\cal H} = 4 \frac{G_{l'l}}{\sqrt{2}}
\sum_{\epsilon,\omega = R,L}^{n = S,V,T}
g^n_{l'_\epsilon l^{\phantom{'}}_\omega}
\left[ \overline{l'_\epsilon}
\Gamma^n {(\nu_{l'})}_\sigma \right]\,
\left[ \overline{({\nu_l})_\lambda} \Gamma_n
	l_\omega \right]\ .
\label{eq:hamiltonian}
\ee
The label $n$ refers to the type of interaction, namely
\be
\Gamma^S = 1\ ,
\hspace{10mm}
\Gamma^V = \gamma^\mu\ ,
\hspace{10mm}
\Gamma^T = \frac{1}{\sqrt{2}} \,\sigma^{\mu \nu}
\equiv \frac{i}{2 \sqrt{2}} \,
( \gamma^\mu \gamma^\nu - \gamma^\nu \gamma^\mu)\ ,
\ee
for the scalar, vector and tensor interactions, respectively.
The neutrino chiralities, $\sigma$ and $\lambda$, are uniquely determined
once $n$ and the charged-lepton chiralities, $\epsilon$ and
$\omega$, are chosen.
Thus, one has 19 real constants, since there are only two non-zero
tensor terms and one global phase may be taken away.

In any reasonable model, these couplings are the low-energy limit
of scalar and/or vector-boson mediated transitions.
In general,
several such contributions will exist and we write
\be
g^n_{l'_\epsilon l^{\phantom{'}}_\omega} =
w^n_{l'_\epsilon l^{\phantom{'}}_\omega} +
a^n_{l'_\epsilon l^{\phantom{'}}_\omega} +
b^n_{l'_\epsilon l^{\phantom{'}}_\omega} +
\cdots\ \ ,
\ee
where the letter $w$ is reserved for the known $W$ boson,
and each letter ($a$, $b$, \ldots) refers to couplings originating
from a given additional intermediate boson.
{}From $W$ decays, as well as from $\beta$ and $\mu$ decays,
we know that the $W$ vertices with leptons will
necessarily give the dominant contribution to the $\tau$ and
$\mu$ leptonic decays.
The SM predicts that this is the only contribution, and,
moreover, that there are only couplings to left-handed leptons.
Hence, in the SM
\be
g^V_{l'_L l^{\phantom{'}}_L} \equiv w^V_{l'_L l^{\phantom{'}}_L} = 1\ ,
\ee
and all other couplings are predicted to vanish.
Of course, what one measures in these decays are the sum
$g^n_{l'_\epsilon l^{\phantom{'}}_\omega}$ of all the different contributions
with the same chiral structure and these may interfere
constructively or destructively.

For an initial lepton-polarization ${\cal P}_l$,
the final charged lepton distribution in the decaying lepton rest frame
is usually parametrized in the form  \cite{BM:57,KS:57}
\begin{eqnarray}\label{eq:spectrum}
{d^2\Gamma(x,\cos\theta) \over dx\, d\cos\theta} &\!\!\! = &\!\!\!
{m_l\omega^4 \over 2\pi^3}(G_{l'l}^2 N) \sqrt{x^2-x_0^2}\,
  \Biggl\{ x (1 - x) + {2\over 9} \rho
 \left(4 x^2 - 3 x - x_0^2 \right)
+  \eta\, x_0 (1-x)
\Biggr.\nonumber\\ & & \Biggl.
  - {1\over 3}{\cal P}_l \, \xi \, \sqrt{x^2-x_0^2} \cos{\theta}
  \left[ 1 - x + {2\over 3}  \delta \left( 4 x - 4 + \sqrt{1-x_0^2}
\right)\right]  \Biggr\} \, , \quad
\end{eqnarray}
where $\theta$ is the angle between the $l^-$ spin and the
final charged-lepton momentum,
$\, \omega \equiv (m_l^2 + m_{l'}^2)/2 m_l \, $
is the maximum $l'^-$ energy for massless neutrinos, $x \equiv E_{l'^-} /
\omega$ is the reduced energy and $x_0\equiv m_{l'}/\omega$.
For unpolarized $l's$, the distribution is characterized by
the so-called Michel \cite{MI:50} parameter $\rho$
and the low-energy parameter $\eta$. Two more parameters, $\xi$
and $\delta$ can be determined when the initial lepton polarization is known.
If the polarization of the final charged lepton is also measured,
5 additional independent parameters \cite{PDG:94}
($\xi'$, $\xi''$, $\eta''$, $\alpha'$, $\beta'$)
appear.

To determine the constraints on
Physics Beyond the SM, it is convenient to express the
Michel parameters in terms of their deviation from
the SM values \cite{mursula}. One obtains,
\ba
\rho - \frac{3}{4}
 & = &
- \frac{3}{4 N}
\left[ {|g^V_{LR}|}^2 + {|g^V_{RL}|}^2 + 2 {|g^T_{LR}|}^2
+ 2{|g^T_{RL}|}^2 +
\mbox{\rm Re}(g^S_{LR} g^{T \ast}_{LR} + g^S_{RL} g^{T \ast}_{RL})
\right]\ ,
\nonumber\\
\eta
& = &
\frac{1}{2 N}
\mbox{\rm Re}\left[
g^S_{LL} g^{V \ast}_{RR} + g^S_{LR} g^{V \ast}_{RL}
+ g^S_{RL} g^{V \ast}_{LR} +g^S_{RR} g^{V \ast}_{LL}
+ 6(g^V_{LR} g^{T \ast}_{RL} +g^V_{RL} g^{T \ast}_{LR})
\right]\ ,
\nonumber\\
\xi - 1
 & = &
- \frac{1}{2 N}
\left[ {|g^S_{LR}|}^2 + {|g^S_{RR}|}^2
+ 4 (-{|g^V_{LR}|}^2 + 2 {|g^V_{RL}|}^2 + {|g^V_{RR}|}^2)
\right.
\nonumber\\
  &  &
\hspace{10mm}
\left.
- 4 {|g^T_{LR}|}^2 + 16 {|g^T_{RL}|}^2
- 8 {\rm Re}(g^S_{LR} g^{T \ast}_{LR} - g^S_{RL} g^{T \ast}_{RL})
\right]\ ,
\\
({\xi}\delta) - \frac{3}{4}
& = &
- \frac{3}{4 N}
\left[ \frac{1}{2} ({|g^S_{LR}|}^2 + {|g^S_{RR}|}^2)
+ ({|g^V_{LR}|}^2 + {|g^V_{RL}|}^2 + 2 {|g^V_{RR}|}^2)
\right.
\nonumber\\
&   &
\hspace{10mm}
\left.
+ 2 ({2 |g^T_{LR}|}^2 + {|g^T_{RL}|}^2)
- \mbox{\rm Re}(g^S_{LR} g^{T \ast}_{LR} - g^S_{RL} g^{T \ast}_{RL})
\right]\ . \nonumber
\label{eq:michel}
\ea
We set the overall normalization factor
\ba
N
& \equiv &
\frac{1}{4} ({|g^S_{LL}|}^2 + {|g^S_{LR}|}^2
+ {|g^S_{RL}|}^2 + {|g^S_{RR}|}^2)
+ ({|g^V_{LL}|}^2 +{|g^V_{LR}|}^2 + {|g^V_{RL}|}^2 + {|g^V_{RR}|}^2)
\nonumber\\
  &  &
+ 3 ({|g^T_{LR}|}^2 + {|g^T_{RL}|}^2)\ ,
\label{eq:N}
\ea
to 1, as it is frequently done.
This may always be done\footnote{
%
%%%%%%%%%%%%%%%%% BEGIN FOOTNOTE %%%%%%%%%%%%%%%%%%%%%%%%%
%
Alternatively, one may absorb $N$ for the
($\mu$,$e$) pair, say, use a common $G_F$ in
Eq.~\ref{eq:hamiltonian}, and keep the normalization
factor $N$ for the other two decays.
However, care must then be taken when using published
bounds for the coupling constants
$g^n_{l'_\epsilon l^{\phantom{'}}_\omega}$,
since the normalization in Eq.~(\ref{eq:N})
is usually adopted.
%often used in deriving those bounds.
}
%
%%%%%%%%%%%%%%%%% END FOOTNOTE %%%%%%%%%%%%%%%%%%%%%%%%%
%
by absorbing it
in the definition of $G^2_{l'l}$.
We note that the parameters
$\eta$ and $G^2_{l'l}$ are the only ones
linear in the new-physics contributions.
Namely, they have terms proportional to
\be
\eta \sim \frac{1}{2} \mbox{\rm Re}(1 \times g^{S \ast}_{RR})\ ,
\label{eq:lineareta}
\ee
and
\be
G^2_{l'l} \propto
1 + 2 \mbox{\rm Re}(1 \times \Delta g^{V \ast}_{LL})\ ,
\label{eq:linearN}
\ee
where we have used the fact that the SM contribution to $g^V_{LL}$
is approximately 1,
and new contributions to $g^V_{LL}$ have been
parametrized by $\Delta g^V_{LL}$.
Clearly this last type of variation is only detectable if it
is non-universal.

It is convenient to introduce \cite{FGJ:86} the probabilities
$Q_{\epsilon\omega}$ for the
decay of a $\omega$-handed $l^-$
into an $\epsilon$-handed
daughter lepton,
\begin{eqnarray}\label{eq:Q_LL}
Q_{LL} &\!\!\! = &\!\!\!
{1 \over 4} |g^S_{LL}|^2 \, + \, |g^V_{LL}|^2
\;\; \phantom{+ \, 3 |g^T_{LR}|^2}
 = {1 \over 4}\left(
-3 +{16\over 3}\rho -{1\over 3}\xi +{16\over 9}\xi\delta +\xi'+\xi''
\right) , \qquad\\ \label{eq:Q_RR}
Q_{RR} &\!\!\! = &\!\!\!
{1 \over 4} |g^S_{RR}|^2 \, + \, |g^V_{RR}|^2
\; \phantom{+ \, 3 |g^T_{LR}|^2}
 =  {1 \over 4}\left(
-3 +{16\over 3}\rho +{1\over 3}\xi -{16\over 9}\xi\delta -\xi'+\xi''
\right) , \quad\\ \label{eq:Q_LR}
Q_{LR} &\!\!\! = &\!\!\!
{1 \over 4} |g^S_{LR}|^2 \, + \, |g^V_{LR}|^2
           \, + \, 3 |g^T_{LR}|^2
 = {1 \over 4}\left(
5 -{16\over 3}\rho +{1\over 3}\xi -{16\over 9}\xi\delta +\xi'-\xi''
\right) , \quad\;\\ \label{eq:Q_RL}
Q_{RL} &\!\!\! = &\!\!\!
{1 \over 4} |g^S_{RL}|^2 \, + \, |g^V_{RL}|^2
           \, + \, 3 |g^T_{RL}|^2
= {1 \over 4}\left(
5 -{16\over 3}\rho -{1\over 3}\xi +{16\over 9}\xi\delta -\xi'-\xi''
\right) . \quad\;
\end{eqnarray}
Upper bounds on any of these (positive-semidefinite) probabilities translate
into corresponding limits for all couplings with the given chiralities.

The total decay rate is given by
\be\label{eq:gamma}
\Gamma\, = \, {m_l^5 G_{l'l}^2\over 192 \pi^3}\,
\left\{ f\!\left({m_{l'}^2\over m_l^2}\right)
+ 4\eta\, {m_{l'}\over m_l}\, g\!\left({m_{l'}^2\over m_l^2}\right)
\right\} r_{\mbox{\rms RC}}
\, ,
\ee
where
\begin{eqnarray}\label{eq:f_g}
f(z) &\!\!\! = &\!\!\! 1 - 8 z + 8 z^3 - z^4 - 12 z^2 \ln{z} \, , \\
g(z) &\!\!\! = &\!\!\! 1 + 9 z - 9 z^2 - z^3 + 6 z (1+z) \ln{z} \, .
\end{eqnarray}
Thus, the normalization $G_{e \mu}$ corresponds to the Fermi coupling $G_F$,
measured in $\mu$ decay.

The factor
\be\label{eq:r_RC}
r_{\mbox{\rms RC}}
 \, = \, \left[1 + {\alpha(m_l) \over 2 \pi }
\left({25 \over 4}  - \pi^2 \right) \right] \,
\left[ 1 + { 3 \over 5 } {m_l^2 \over M_W^2}
- 2 {m_{l'}^2 \over M_W^2}\right]
\, ,
\ee
takes into account radiative corrections not
included in the
Fermi coupling constant $G_F$, and the non-local structure of the
$W$ propagator. These effects \cite{MS:88} are quite small:
%$r_{\mbox{\rms RC}}[\tau\to\mu,e] = 0.9960$;
%$r_{\mbox{\rms RC}}[\mu\to e] = 0.9958$.
$r_{\mbox{\rms RC}}^{\tau\to\mu,e} = 0.9960$;
$r_{\mbox{\rms RC}}^{\mu\to e} = 0.9958$.
Notice, that the we are adopting the usual procedure of taking the
radiative corrections within the Standard Model.
Since we assume that the Standard Model provides the dominant contribution
to the decay rate, any additional higher-order correction
beyond the effective four-fermion Hamiltonian (\ref{eq:hamiltonian})
would be a subleading effect.

The kinematical integrations have been done assuming massless neutrinos.
The numerical correction induced by a non-zero $\nu_l$ mass,
$r_{\nu_l} \equiv 1 + \delta_{\nu_l}\approx 1 - 8 (m_{\nu_l}/m_l)^2$,
is quite small. The present experimental upper limits \cite{PDG:94,ALEPH:95b}
on the neutrino masses imply:
$|\delta_{\nu_\mu}^{\mu\to e}| < 5 \times 10^{-5}$ (90\% CL),
$|\delta_{\nu_\mu}^{\tau\to\mu}| < 2 \times 10^{-7}$ (90\% CL),
$|\delta_{\nu_\tau}^{\tau\to\mu,e}| < 1.4 \times 10^{-3}$ (95\% CL).

It is fortunate that the two parameters which are linear in the
new-physics contributions,
$\eta$ and $G^2_{l'l}$,
are precisely the ones which
survive in the total decay width. One can then study them
with non-universality searches which already provide very precise tests
of the lepton sector.

\section{Experimental summary}
\label{sec:exp}

For $\mu$-decay, where precise measurements of the polarizations of
both $\mu$ and $e$ have been performed, there exist \cite{FGJ:86}
upper bounds on $Q_{RR}$, $Q_{LR}$ and $Q_{RL}$, and a lower bound
on $Q_{LL}$. They imply corresponding upper bounds on the 8
couplings $|g^n_{RR}|$, $|g^n_{LR}|$ and $|g^n_{RL}|$.
The measurements of the $\mu^-$ and the $e^-$ do not allow us to
determine $|g^S_{LL}|$ and $|g^V_{LL}|$ separately \cite{FGJ:86,JA:66}.
Nevertheless, since the helicity of the $\nu_\mu$ in pion decay is
experimentally known to be $-1$, a lower limit on $|g^V_{LL}|$ is
obtained \cite{FGJ:86} from the inverse muon decay
$\nu_\mu e^-\to\mu^-\nu_e$.
The present (90\% CL) bounds \cite{PDG:94} on the $\mu$-decay couplings
are given in Table~\ref{tab:mu_couplings}. These limits show nicely
that the bulk of the $\mu$-decay transition amplitude is indeed of
the predicted V$-$A type.

%%%%%%%%%%%%  TABLE  %%%%%%%%%%%%
\begin{table}[hbt]
\centering
\begin{tabular}{|l|l|l|}
\hline
$|g^S_{e_L \mu_L}| < 0.55$  & $|g^V_{e_L \mu_L}| > 0.96$ &
\hfil -- \hfil \\
$|g^S_{e_R \mu_R}| < 0.066$ & $|g^V_{e_R \mu_R}| < 0.033$ &
\hfil -- \hfil \\
$|g^S_{e_L \mu_R}| < 0.125$ & $|g^V_{e_L \mu_R}| < 0.060$ &
 $|g^T_{e_L \mu_R}| < 0.036$\\
$|g^S_{e_R \mu_L}| < 0.424$ & $|g^V_{e_R \mu_L}| < 0.110$ &
 $|g^T_{e_R \mu_L}| < 0.122$\\
\hline
\end{tabular}
\caption{90\% CL experimental limits \protect\cite{PDG:94}
for the $\mu$-decay $g^n_{e_\epsilon \mu_\omega}$ couplings.}
\label{tab:mu_couplings}
\end{table}
%%%%%%%%%%%%%%%%%%%%%%%%%%%%%%%%%

The experimental analysis of the $\tau$-decay parameters is necessarily
different from the one applied to the muon, because of the much
shorter $\tau$ lifetime.
The measurement of the $\tau$ polarization and the parameters
$\xi$ and $\delta$ is still possible due to the fact that the spins
of the $\tau^+\tau^-$ pair produced in $e^+e^-$ annihilation
are strongly correlated [14--23].
%\cite{TS:71,KST:73,PS:77,GO:89,NE:91,GN:91,FE:90,BPR:91,ABGPR:92,DDDR:93}.
However,
the polarization of the charged lepton emitted in the $\tau$ decay
has never been measured. In principle, this could be done
for the decay $\tau^-\to\mu^-\bar\nu_\mu\nu_\tau$ by stopping the
muons and detecting their decay products \cite{FE:90}.
The measurement of the inverse decay $\nu_\tau l^-\to\tau^-\nu_l$
looks far out of reach.

The present experimental status on the $\tau$-decay Michel parameters
is shown in Table~\ref{tab:tau_michel}, which gives the world-averages
of all published
\cite{ALEPH:95,ARGUS:95,ARGUS:95b,PDG:94}
measurements.
The improved accuracy of the most recent experimental analyses
has brought an enhanced sensitivity to the different shape parameters,
allowing the first measurements of $\eta_{\tau\to\mu}$
\cite{ALEPH:95,ARGUS:95},
$\xi_{\tau\to e}$, $\xi_{\tau\to\mu}$, $(\xi\delta)_{\tau\to e}$ and
$(\xi\delta)_{\tau\to\mu}$ \cite{ALEPH:95}.
(The ARGUS measurement \cite{ARGUS:95b} of $\xi_{\tau\to l}$ and
$(\xi\delta)_{\tau\to l}$ assumes identical couplings for $l=e,\mu$.
A measurement of $\sqrt{\xi_{\tau\to e}\xi_{\tau\to\mu}}$
was published previously \cite{ARGUS:93}).

%%%%%%%%%%%%%%% Table %%%%%%%%%%%%%
\begin{table}[htb]
\centering
\begin{tabular}{|c|c|c|c|c|}
\hline
Parameter & $\tau^-\to\mu^-$ & $\tau^-\to e^-$ &
With Lepton-Universality & SM
\\ \hline
$\rho$ & $0.738\pm 0.038$ & $0.736\pm 0.028$ & $0.733\pm 0.022$ & 0.75
\\
$\eta$ & $-0.14\pm 0.23\phantom{-}$ & -- & $-0.01\pm 0.14\phantom{-}$ & 0
\\
$\xi$ & $1.23\pm 0.24$ & $1.03\pm 0.25$ & $1.06\pm 0.11$ & 1
\\
$\xi\delta$ & $0.71\pm 0.15$ & $ 1.11\pm 0.18$ & $ 0.76\pm 0.09$ & 0.75
\\ \hline
\end{tabular}
\caption{Experimental averages of the $\tau$-decay Michel parameters
\protect\cite{ALEPH:95,ARGUS:95,ARGUS:95b,PDG:94}.
The fourth column assumes lepton universality.}
\label{tab:tau_michel}
\end{table}
%%%%%%%%%%%%%%%%%%%%%%%%%%%%%%%%%

The determination of the $\tau$-polarization
parameters \cite{ALEPH:95,ARGUS:95b,RO:95},
allows us to bound the total probability for the decay of
a right-handed $\tau$,
\be\label{eq:Q_R}
Q_{\tau_R} \equiv Q_{l'_R\tau^{\phantom{'}}_R} + Q_{l'_L\tau^{\phantom{'}}_R}
= \frac{1}{2}\, \left[ 1 + \frac{\xi}{3} - \frac{16}{9} (\xi\delta)\right]
\; .
\ee
One finds (ignoring possible correlations among the measurements):
\begin{eqnarray}
Q_{\tau_R}^{\tau\to\mu} &\!\!\! =&\!\!\! \phantom{-}0.07\pm 0.14 \;
< \, 0.28 \quad (90\%\;\mbox{\rm CL})\, , \\
Q_{\tau_R}^{\tau\to e} &\!\!\! =&\!\!\! -0.32\pm 0.17 \;
< \, 0.14 \quad (90\%\;\mbox{\rm CL})\, , \\
Q_{\tau_R}^{\tau\to l} &\!\!\! =&\!\!\! \phantom{-}0.00\pm 0.08 \;
< \, 0.14 \quad (90\%\;\mbox{\rm CL})\, ,
\end{eqnarray}
where the last value refers to the $\tau$-decay into either $l=e$ or $\mu$,
assuming universal leptonic couplings.
Since these probabilities are positive semidefinite quantities, they imply
corresponding limits on all $|g^n_{l_R\tau_R}|$
and $|g^n_{l_L\tau_R}|$ couplings.
The quoted 90\% CL have been obtained
adopting a Bayesian approach for one-sided limits \cite{PDG:94}.
Table~\ref{table:g_tau_bounds} gives the implied bounds on the
$\tau$-decay couplings.

%%%%%%%%%%%%  TABLE  %%%%%%%%%%%%
\begin{table}[hbt]
\centering
\begin{tabular}{||l|l||l||}
\hline
\hfil $\tau\to\mu$\hfil &\hfil $\tau\to e$ \hfil &
\hfil $\tau\to l$ \hfil
\\\hline\hline
$|g^S_{\mu_R\tau_R}| < 1.05$ &
$|g^S_{e_R\tau_R}| < 0.75^*$ &
$|g^S_{l_R\tau_R}| < 0.74$
\\
$|g^S_{\mu_L\tau_R}| < 1.05$ &
$|g^S_{e_L\tau_R}| < 0.75^*$ &
$|g^S_{l_L\tau_R}| < 0.74$
\\ \hline
$|g^V_{\mu_R\tau_R}| < 0.53$ &
$|g^V_{e_R\tau_R}| < 0.38^*$ &
$|g^V_{l_R\tau_R}| < 0.37$
\\
$|g^V_{\mu_L\tau_R}| < 0.53$ &
$|g^V_{e_L\tau_R}| < 0.38^*$ &
$|g^V_{l_L\tau_R}| < 0.37$
\\ \hline
$|g^T_{\mu_L\tau_R}| < 0.30$ &
$|g^T_{e_L\tau_R}| < 0.22^*$ &
$|g^T_{l_L\tau_R}| < 0.21$
\\ \hline
\end{tabular}
\caption{90\% CL limits
for the $\tau_R$-decay $g^n_{l_\epsilon \tau_R}$ couplings.
The numbers with an asterisk use the measured value of $(\xi\delta)_e$;
the meaning of the assigned confidence level could be doubtful in this case
(see text).}
\label{table:g_tau_bounds}
\end{table}
%%%%%%%%%%%%%%%%%%%%%%%%%%%%%%%%%

Notice, however, that the central value of $Q_{\tau_R}^{\tau\to e}$ turns out
to be negative at the $2\sigma$ level; i.e.~, there is only a 3\%
probability to have a positive value of
$Q_{\tau_R}^{\tau\to e}$. Therefore, the limits on
$|g^n_{e_R\tau_R}|$ and $|g^n_{e_L\tau_R}|$
should be taken with some caution, since the meaning of the assigned
confidence level is not at all clear.

The problem clearly comes from the measured value of $(\xi\delta)_e$.
In order to get a positive probability
$Q_{\tau_R}$, one needs
$(\xi -1) > \frac{16}{3} [(\xi\delta) -\frac{3}{4}]$.
Thus, $(\xi\delta)$ can only be made larger than
$3/4$ at the expense of making $\xi$ correspondingly much
larger than one.
Hence, if the current values of the Michel parameters
for the decay of tau into electron and neutrinos were to be
confirmed, one would have to go beyond the effective hamiltonian
of Eq.~(\ref{eq:hamiltonian}):
{\it the combined observations for $\xi_{\tau\to e}$ and
$(\xi\delta)_{\tau\to e}$ are
not consistent with an effective four-fermion interaction of
the form in Eq.~(\ref{eq:hamiltonian})}. That is to say that
no flavour-conserving, derivative-free, four-lepton interaction can
be found, satisfying both these results simultaneously.
Further, since lepton-flavour violations have no measurable
effect if the final neutrinos are massless and unobserved
\cite{langacker},
and derivative couplings would be suppressed by
$m_\tau^2/M_W^2 \sim 5 \times 10^{-4}$, a sizeable effect not included
in the effective Hamiltonian (\ref{eq:hamiltonian}) seems very
unlikely\footnote{
%%%%%%%
The alternative is to go beyond the four-fermion Hamiltonian
(\protect\ref{eq:hamiltonian}), allowing, for example, the decay
of the tau into an electron and two (unobserved) neutral scalars,
such as Majorons \protect\cite{SBP:87} or supersymmetric
scalar neutrinos.}.
%Although a general analysis does not exist, the inequality
%$(\xi-1)> (\xi\delta-\frac{3}{4})$ is also satisfied in the Majoron
%model studied in Ref.~\protect\cite{SBP:87}.}.   ?????
%%%%%%%%%%%
Hence, based solely on theoretical grounds, one can
achieve the conclusion that, within a four-fermion hamiltonian,
either $(\xi\delta)_{\tau\to e}$ comes into agreement with the SM,
or $\xi_{\tau\to e}$ must move by a factor close to $2$
(a most unreasonable proposition).

Table~\ref{tab:parameters} gives the world-average values of
$m_\tau$, $\tau_\tau$,
$B_l\equiv\mbox{\rm Br}(\tau^-\to\nu_\tau l^-\bar\nu_l)$,
$B_\pi\equiv\mbox{\rm Br}(\tau^-\to\nu_\tau\pi^-)$,
$B_K\equiv\mbox{\rm Br}(\tau^-\to\nu_\tau K^-)$ and
$B_h\equiv\mbox{\rm Br}(\tau^-\to\nu_\tau\pi^- + \nu_\tau K^-)$.
In view of the significant improvements achieved with the most recent data,
updated numbers including preliminary results reported in the last
$\tau$ Workshop \cite{MO:94} are also given.

%%%%%%%%%%%%%%%%%%%%%%%%%% Table %%%%%%%%%%%%%%%%%%%%%%%%%%%%%%%%%
%
\begin{table}[thb]
\centering
\begin{tabular}{|c|c|c|}
\hline
Parameter & PDG 94 & Montreux 94 \\ \hline
$m_\tau$ & $(1777.1^{+0.4}_{-0.5})$ MeV & $(1777.0\pm 0.3)$ MeV \\
$\tau_\tau$ & $(295.6\pm 3.1)\times 10^{-15}$ s &
$(291.6\pm 1.6)\times 10^{-15}$ s \\
$B_e$ & $(18.01\pm 0.18)\% $ & $(17.79\pm 0.09)\% $ \\
$B_\mu$ & $(17.65\pm 0.24)\% $ & $(17.33\pm 0.09)\% $
\\
$B_\pi$ & $(11.7\pm 0.4)\% $ & $(11.09\pm 0.15)\% $ \\
$B_K$ & $(0.67\pm 0.23)\% $ & $(0.68\pm 0.04)\% $ \\
$B_h$ & $(12.88\pm 0.34)\% $ & $(11.77\pm 0.14)\% $
\\ \hline
\end{tabular}
\caption{World-average values \protect\cite{PDG:94} of the
$\tau$ mass, lifetime, leptonic branching ratios and
$\mbox{\rm Br}(\tau^-\to\nu_\tau\pi^-/K^-)$.
The updated numbers in the third column, include
{\it preliminary} results reported in the last $\tau$ Workshop
\protect\cite{MO:94}.}
\label{tab:parameters}
\end{table}
%
%%%%%%%%%%%%%%%%%%%%%%%%%%%%%%%%%%%%%%%%%%%%%%%%%%%%%%%%%%%%%%%%%%%%%

\section{Universality tests}
\label{sec:universality}

The universality of the leptonic couplings can be tested through
the ratios of the measured leptonic-decay widths:
\begin{eqnarray} \label{eq:Gmu/Ge}
{\Gamma_{\tau\to\mu}\over\Gamma_{\tau\to e}}
& \Longrightarrow & \left|{\widehat G_{\mu\tau}
%\sqrt{1 + 4\eta_{\tau\to\mu} \left({m_\mu\over m_\tau}\right)}
\over \widehat G_{e\tau}
%\sqrt{1 + 4\eta_{\tau\to e} \left({m_e\over m_\tau}\right)}
}\right|
= \left\{
  \begin{array}{cc} 1.0038\pm 0.0087 \quad & \mbox{\rm (PDG 94)} \\
  1.0008 \pm 0.0036 \quad & \mbox{\rm (Montreux 94)}
  \end{array}\right. ,
\\  \label{eq:Gtau/Gmu}
{\Gamma_{\tau\to\mu}\over\Gamma_{\mu\to e}}
& \Longrightarrow & \left|{\widehat G_{\mu\tau}
%\sqrt{1 + 4\eta_{\tau\to\mu} \left({m_\mu\over m_\tau}\right)}
\over \widehat G_{e\mu}
%\sqrt{1 + 4\eta_{\mu\to e} \left({m_e\over m_\mu}\right)}
}\right|
=
\left\{ \begin{array}{cc}
0.9970\pm 0.0073 \quad &\mbox{\rm (PDG 94)} \\
  0.9979 \pm 0.0037 \quad & \mbox{\rm (Montreux 94)}
\end{array}\right. ,
\end{eqnarray}
where
\be\label{eq:Ghat_def}
\widehat G_{l'l} \,\equiv\, G_{l'l} \,
\sqrt{1 + 4\,\eta_{\l\to l'}\, {m_{l'}\over m_l}\,
{g\!\left( m_{l'}^2/ m_l^2 \right)\over  f\!\left( m_{l'}^2/ m_l^2 \right)}}
\, .
\ee

An important point, emphatically stressed by
Fetscher and Gerber \cite{fgreview}, concerns the extraction
of $G_{e \mu}$ from $\mu$ decays, whose uncertainty is dominated
by the uncertainty in $\eta$.

In models where $\eta=0$, $\widehat G_{l'l} = G_{l'l}$; then
the limits (\ref{eq:Gmu/Ge}) and
(\ref{eq:Gtau/Gmu})
strongly constrain
possible deviations from universality.
To first-order in new physics,
$G_{l'l}\propto 1 + \mbox{\rm Re}(\Delta g_{LL}^V)$.
Therefore, at 90\% CL,
$-0.005 \ (-0.010) <
\mbox{\rm Re}(\Delta g_{\mu_L\tau_L}^V- \Delta g_{e_L\tau_L}^V) < 0.007$
(0.018) and
$-0.008 \ (-0.015) <
\mbox{\rm Re}(\Delta g_{\mu_L\tau_L}^V- \Delta g_{e_L\mu_L}^V) < 0.004$
(0.009), using the Montreux 94  (PDG~94) data.

Conversely, if lepton universality is assumed (i.e. $G_{l'l} = G_F$,
$\, g^n_{l'_\epsilon l^{\phantom{'}}_\omega} \! = g^n_{\epsilon\omega}$),
the leptonic decay ratios (\ref{eq:Gmu/Ge})  and (\ref{eq:Gtau/Gmu})
provide limits on the low-energy parameter $\eta$.
The best sensitivity \cite{stahl} comes from
$\widehat G_{\mu\tau}$,
where the term proportional to $\eta$ is not suppressed by
the small $m_e/m_l$ factor. The measured $B_\mu/B_e$ ratio implies
then:
\be\label{eq:eta_univ}
\eta \, = \, \left\{
\begin{array}{cc} 0.034 \pm 0.076 \quad &\mbox{\rm (PDG 94)} \\
  0.007\pm 0.033    \quad & \mbox{\rm (Montreux 94)}
\end{array}\right. .
\ee
This determination is more accurate that the one in
Table~\ref{tab:tau_michel},
obtained from the shape of the energy distribution,
and is comparable to the value measured in $\mu$-decay:
$\eta_{\mu\to e} = -0.007\pm 0.013$ \cite{PDG:94}.

A non-zero value of $\eta$ would show that there are at least two
different couplings with opposite chiralities for the charged leptons.
Since, we assume the V$-$A coupling $g_{LL}^V$ to be dominant, the
second coupling would be \cite{FE:90} a Higgs type coupling $g^S_{RR}$
[$\eta\approx\mbox{\rm Re}(g^S_{RR})/2$,
to first-order in new-physics contributions].
Thus, Eq.~(\ref{eq:eta_univ}) puts the (90\% CL) bound:
$-0.09 \, (-0.18) <\mbox{\rm Re}(g^S_{RR}) < 0.12$ (0.32),
using the Montreux 94 (PDG 94) data.

Finally, in models in which the new-physics couples exclusively to
the lepton sector (so that the CKM matrix is unitary),
further information may be found by comparing  $G_{l'l}$
with $G_F$ as extracted from the combination
of $\beta$ and $K_{e3}$ decays \cite{barroso}.
Indeed, the usually quoted values for the CKM angles are extracted
assuming that the coupling constant $g$, coupling $W$ to fermions,
is the same for quarks and leptons.
Thus, if there are new contributions affecting only the lepton couplings,
any deviation from unitarity in the first row of the CKM matrix
reflects a deviation of $g_\mu$ from the SM value.

\subsection{$W$-exchange model}

The universality constraints are commonly presented,
assuming that the leptonic decays proceed exclusively
through the SM V$-$A interaction.
In that case the $\widehat G_{l'l}$ ratios reduce
to the corresponding ratios of leptonic $W$-couplings:
$|\widehat G_{\mu\tau}/\widehat G_{e\tau}| = |g_\mu/g_e|$;
$|\widehat G_{\mu\tau}/\widehat G_{e\mu}| = |g_\tau/g_e|$.
Eq.~(\ref{eq:Gmu/Ge}) should then be compared
with the more accurate value \cite{BR:92,CZ:93}
\be\label{eq:univ_pi}
\left| {g_{\mu} \over g_e} \right| = 1.0017 \pm 0.0015 \, ,
\ee
obtained from the ratio
$R_{e/\mu}\equiv\Gamma(\pi^-\to e^-\bar\nu_e)/
\Gamma(\pi^-\to\mu^-\bar\nu_\mu)$.

The decay modes $\tau^-\to\nu_\tau\pi^-$ and $\tau^-\to\nu_\tau K^-$
can also be used to test universality through the ratios
\begin{eqnarray}\label{eq:R_tp}
R_{\tau/\pi} & \!\!\!\equiv &\!\!
 {\Gamma(\tau^-\to\nu_\tau\pi^-) \over
 \Gamma(\pi^-\to \mu^-\bar\nu_\mu)} =
\left| {g_\tau\over g_\mu}\right|^2
{m_\tau^3\over 2 m_\pi m_\mu^2} \,
{(1-m_\pi^2/ m_\tau^2)^2\over
 (1-m_\mu^2/ m_\pi^2)^2}
\left( 1 + \delta R_{\tau/\pi}\right) , \qquad
\\ \label{eq:R_tk}
R_{\tau/K} &\!\!\! \equiv &\!\!\! {\Gamma(\tau^-\to\nu_\tau K^-) \over
 \Gamma(K^-\to \mu^-\bar\nu_\mu)} =
\left| {g_\tau\over g_\mu}\right|^2
{m_\tau^3\over 2 m_K m_\mu^2}
{(1-m_K^2/m_\tau^2)^2\over
(1-m_\mu^2/ m_K^2)^2}
\left( 1 + \delta R_{\tau/K}\right) , \qquad
\end{eqnarray}
where the dependence on the hadronic matrix elements (the so-called
decay constants $f_{\pi,K}$) factors out.
Owing to the different energy scales involved, the radiative
corrections to the $\tau^-\to\nu_\tau\pi^-/K^-$ amplitudes
are however not the same than the corresponding effects in
$\pi^-/K^-\to\mu^-\bar\nu_\mu$. The size of the relative
correction has been estimated by
Marciano and Sirlin \cite{MS:93} to be
$\delta R_{\tau/\pi} = (0.67\pm 1.)\% $, where the 1\% error
is due to the missing long-distance contributions to the
tau decay rate.
A recent evaluation of those long-distance corrections \cite{DF:94}
quotes the more precise values
\be\label{eq:dR_tp_tk}
\delta R_{\tau/\pi} = (0.16\pm 0.14)\% , \qquad\qquad
\delta R_{\tau/K} = (0.90\pm 0.22)\% .
\ee
Using these numbers, the measured $\tau^-\to\pi^-\nu_\tau$
and $\tau^-\to K^-\nu_\tau$ decay rates imply
\be\label{eq:g_tau_mu/pi_k}
\left| {g_\tau\over g_\mu}\right|_\pi = \left\{
\begin{array}{c} 1.027\pm 0.018  \\
  1.006 \pm 0.008 \end{array}\right. ;
\quad
\left| {g_\tau\over g_\mu}\right|_K = \left\{
\begin{array}{cc} 0.96\pm 0.17 \quad & \mbox{\rm (PDG 94)} \\
  0.972\pm 0.029 \quad & \mbox{\rm (Montreux 94)} \end{array}\right. .
\ee
The inclusive sum of both decay modes, i.e.
$\Gamma[\tau^-\to h^-\nu_\tau]$ with $h=\pi,K$, provides a
slightly more accurate determination:
\be\label{eq:g_tau_mu/pi/k}
 \left| {g_\tau\over g_\mu}\right|_{\pi/K} = \left\{
\begin{array}{cc} 1.043\pm 0.015 \qquad & \mbox{\rm (PDG 94)} \\
  1.004\pm 0.007 \qquad & \mbox{\rm (Montreux 94)} \end{array}\right. .
\ee

An independent test of lepton universality has been obtained
at the $p$-$\bar p$ colliders, by comparing the ratios of the
$\sigma \cdot B$  partial production cross-sections
for the various $\, W^- \to l^- \bar\nu_l \, $  decay modes. The
results  of these analyses \cite{UA1:89,UA2:91,CDF:92}
are however less precise:
\begin{equation}
\label{eq:univIV}
\left| {g_{\mu} \over g_e } \right|  =
   1.00 \pm 0.08  \, , \quad
   \left| {g_{\tau} \over g_e } \right|  =
   0.99 \pm 0.04 \,  .
\end{equation}

Thus, the present data verify the universality of the leptonic
charged-current couplings to the 0.16\% ($e/\mu$) and 0.37\%
($\tau/\mu$) level. The precision of the most recent
$\tau$-decay measurements is becoming competitive with the
more accurate $\pi$-decay determination.
It is important to realize the complementarity of the
different universality tests.
The pure leptonic decay modes probe
the charged-current couplings of a transverse $W$. In contrast,
the decays $\pi/K\to l\bar\nu$ and $\tau\to\nu_\tau\pi/K$ are only
sensitive to the longitudinal $W$ couplings.
One can easily imagine new-physics scenarios which would modify
differently the two types of leptonic couplings \cite{MA:94}.
For instance,
in the usual two-Higgs doublet model,
the charged-scalar exchange generates a correction to the ratio
$B_\mu/B_e$,
but the pion-decay ratio $R_{e/\mu}$ remains unaffected.
Similarly, lepton mixing between the $\nu_\tau$ and an hypothetical
heavy neutrino would not modify the ratios  $B_\mu/B_e$ and
$R_{e/\mu}$, but would certainly correct the relation between
$\Gamma(\tau^-\to\nu_\tau l^-\bar\nu_l)$ and
$\Gamma(\mu^-\to\nu_\mu e^-\bar\nu_e)$.

\section{Constraints on new charged bosons}

In this section we assume that the interactions are mediated by
charged vectors and/or charged scalars; therefore, there are no tensor
couplings and Eqs.~(\ref{eq:michel}) become simpler.
In particular,
the quantities $(1-\frac{4}{3}\rho)$ and
$(1-\frac{4}{3}\xi\delta)$ reduce to sums of $|g^n_{l'_\epsilon l_\omega}|^2$,
which are positive semidefinite;
i.e.~,
in the absence of tensor couplings, $\rho\leq\frac{3}{4}$ and
$\xi\delta\leq\frac{3}{4}$.
This allows us to extract direct bounds on
several couplings.

The measured values of $\rho_{\mu\to e}$,
$\rho_{\tau\to\mu}$, $\rho_{\tau\to e}$ and
$\rho_{\tau\to l}$ ($l=e,\mu$) imply:
\be\label{eq:rho_bounds_CH}
\begin{array}{cccc}
|g^V_{e_L\mu_R}|^2 + |g^V_{e_R\mu_L}|^2 \, &=\, -0.0024 \pm 0.0035 \;
&< \; 0.0045 \quad & (90\%\;\mbox{\rm  CL})\, ,
\\
|g^V_{\mu_L\tau_R}|^2 + |g^V_{\mu_R\tau_L}|^2 \, &=\,
\phantom{-0}0.016 \pm 0.051\phantom{0} \; &< \; 0.094 \quad
& (90\%\;\mbox{\rm  CL})\, ,
\\
|g^V_{e_L\tau_R}|^2 + |g^V_{e_R\tau_L}|^2 \, &=\,
\phantom{-0}0.019 \pm 0.037\phantom{0} \; &< \; 0.074\quad
& (90\%\;\mbox{\rm  CL})\, ,
\\
|g^V_{l_L\tau_R}|^2 + |g^V_{l_R\tau_L}|^2 \, &=\,
\phantom{-0}0.023 \pm 0.029\phantom{0} \; &< \; 0.064\quad
& (90\%\;\mbox{\rm  CL})\, .
\end{array}
\ee
Except for  $|g^V_{e_L\mu_R}|$, these limits are
stronger than the general ones in Tables~\ref{tab:mu_couplings} and
\ref{table:g_tau_bounds}.

Similarly, one gets from the different $\xi\delta$ measurements:
\be\label{eq:delta_bounds_CH}
\begin{array}{cl}
\!\!\!\!
|g^V_{e_L\mu_R}|^2 + |g^V_{e_R\mu_L}|^2 + 2 |g^V_{e_R\mu_R}|^2
+ \frac{1}{2} |g^S_{e_L\mu_R}|^2 + \frac{1}{2} |g^S_{e_R\mu_R}|^2
&=\, -0.0017 \pm 0.0096
\\ & < \; 0.015 \quad (90\%\;\mbox{\rm  CL})\, , \quad
\\ \!\!\!\!
|g^V_{\mu_L\tau_R}|^2 + |g^V_{\mu_R\tau_L}|^2 + 2 |g^V_{\mu_R\tau_R}|^2
+ \frac{1}{2} |g^S_{\mu_L\tau_R}|^2 + \frac{1}{2} |g^S_{\mu_R\tau_R}|^2
&=\, 0.05 \pm 0.20
\\ & < \; 0.36 \quad (90\%\;\mbox{\rm  CL})\, ,
\\ \!\!\!\!
|g^V_{e_L\tau_R}|^2 + |g^V_{e_R\tau_L}|^2 + 2 |g^V_{e_R\tau_R}|^2
+ \frac{1}{2} |g^S_{e_L\tau_R}|^2 + \frac{1}{2} |g^S_{e_R\tau_R}|^2
 &=\, -0.48 \pm 0.24
\\ & < \; 0.20 \quad (90\%\;\mbox{\rm  CL})\, ,
\\ \!\!\!\!
|g^V_{l_L\tau_R}|^2 + |g^V_{l_R\tau_L}|^2 + 2 |g^V_{l_R\tau_R}|^2
+ \frac{1}{2} |g^S_{l_L\tau_R}|^2 + \frac{1}{2} |g^S_{l_R\tau_R}|^2
&=\, -0.01 \pm 0.12
\\ & < \; 0.19 \quad (90\%\;\mbox{\rm  CL})\, .
\end{array}
\ee
The limits on the $(\mu,e)$ couplings are weaker than
the ones in Table~\ref{tab:mu_couplings}. The bounds on the vector
LR and RL couplings are also worse than the ones coming from
Eq.~(\ref{eq:rho_bounds_CH}). However, the resulting limits
on the other couplings are  stronger
than the ones in Table~\ref{table:g_tau_bounds}.
The constraint from $(\xi\delta)_{\tau\to e}$ shows explicitly that it is
not possible to accommodate a value larger than $3/4$ with charged-boson
(vector or/and scalar) exchanges.

In the absence of tensor couplings, we can combine the information on $\xi$
and $\rho$ to obtain another positive-semidefinite combination of couplings:
$(1-\frac{4}{3}\rho) + \frac{1}{2} (1-\xi)$. The present data imply:
\be\label{eq:xi_rho}
\begin{array}{cl}
3 |g^V_{e_R\mu_L}|^2 + |g^V_{e_R\mu_R}|^2
+ \frac{1}{4} |g^S_{e_L\mu_R}|^2 + \frac{1}{4} |g^S_{e_R\mu_R}|^2
 &=\,  -0.0039 \pm 0.0053
\\ & < \; 0.0067 \quad (90\%\;\mbox{\rm  CL})\, ,
\\
3 |g^V_{\mu_R\tau_L}|^2 + |g^V_{\mu_R\tau_R}|^2
+ \frac{1}{4} |g^S_{\mu_L\tau_R}|^2 + \frac{1}{4} |g^S_{\mu_R\tau_R}|^2
 &=\,\phantom{00} -0.10 \pm 0.13\phantom{00}
\\ & < \; 0.16 \quad (90\%\;\mbox{\rm  CL})\, ,
\\
3 |g^V_{e_R\tau_L}|^2 + |g^V_{e_R\tau_R}|^2
+ \frac{1}{4} |g^S_{e_L\tau_R}|^2 + \frac{1}{4} |g^S_{e_R\tau_R}|^2
 &=\,\phantom{-00}  0.00 \pm 0.13\phantom{00}
\\ & < \; 0.21 \quad (90\%\;\mbox{\rm  CL})\, ,
\\
3 |g^V_{l_R\tau_L}|^2 + |g^V_{l_R\tau_R}|^2
+ \frac{1}{4} |g^S_{l_L\tau_R}|^2 + \frac{1}{4} |g^S_{l_R\tau_R}|^2
&=\,\phantom{00} -0.01 \pm 0.06\phantom{00}
\\ & < \; 0.10 \quad (90\%\;\mbox{\rm  CL})\, .
\end{array}\ee
The resulting limits on
$|g^V_{e_R\mu_L}|$, $|g^V_{\mu_R\tau_L}|$, $|g^V_{\mu_R\tau_R}|$,
$|g^S_{\mu_L\tau_R}|$, $|g^S_{\mu_R\tau_R}|$, $|g^V_{e_R\tau_L}|$
and $|g^V_{l_R\tau_L}|$ are stronger than the ones obtained before.

Combining the different limits, one gets the bounds shown in
Table~\ref{tab:coup_CH}. The numbers with an asterisk have been
derived from $(\xi\delta)_e$. If this information is not used, one finds
the weaker limits:
$|g^S_{e_R\tau_R}|<0.92$, $|g^S_{e_L\tau_R}|<0.92$ and
$|g^V_{e_R\tau_R}|<0.46$.

%%%%%%%%%%%%  TABLE  %%%%%%%%%%%%
\begin{table}[hbt]
\centering
\begin{tabular}{||l||l|l|l||l||}
\hline & \hfil $\mu\to e$\hfil &
\hfil $\tau\to\mu$\hfil &\hfil $\tau\to e$ \hfil &
\hfil $\tau\to l$ \hfil
\\\hline\hline
$|g^S_{LL}|$ & $<0.55$ & $\leq 2$ & $\leq 2$ & $\leq 2$
\\
$|g^S_{RR}|$ & $<0.066$ & $<0.80$ & $<0.63^*$ & $<0.62$
\\
$|g^S_{LR}|$ & $<0.125$ & $<0.80$ & $<0.63^*$ & $<0.62$
\\
$|g^S_{RL}|$ & $<0.424$ & $\leq 2$ & $\leq 2$ & $\leq 2$
\\ \hline
$|g^V_{LL}|$ & $>0.96$ & $\leq 1$ & $\leq 1$ & $\leq 1$
\\
$|g^V_{RR}|$ & $<0.033$ & $<0.40$ & $<0.32^*$ & $<0.31$
\\
$|g^V_{LR}|$ & $<0.060$ & $<0.31$ & $<0.27$ & $<0.25$
\\
$|g^V_{RL}|$ & $<0.047$ & $<0.23$ & $<0.27$ & $<0.18$
\\ \hline
\end{tabular}
\caption{90\% CL limits
for the couplings $g^n_{\epsilon\omega}$, assuming that there are no
tensor couplings.
The numbers with an asterisk use the measured value of $(\xi\delta)_e$.}
\label{tab:coup_CH}
\end{table}
%%%%%%%%%%%%%%%%%%%%%%%%%%%%%%%%%

Up to now, our only assumption has been the absence of tensor couplings.
However, in many extensions of the SM, the bounds we have derived on the
couplings can be improved due to additional knowledge of
the underlying dynamics.
Such is the case with any model whose deviations from the SM in the
lepton sector are dominated by one intermediate state.
This will typically occur with the least massive gauge boson,
if its couplings are not suppressed by some approximate symmetry.
In the following, we study the constraints associated with
the addition of one `dominant' intermediate boson to the SM.

\subsection{Factorization}

Let us assume that the interactions are mediated by a single
charged boson (either vector or scalar).
Then, the previous limits are improved due to additional relations
among the couplings. Indeed, the factorization
thus implied yields \cite{mursula},
\be
\alpha^n_{LR}\  \alpha^n_{RL} =
\alpha^n_{LL}\  \alpha^n_{RR} \ ,
\label{eq:factorization}
\ee
where we have used $\alpha$
(standing for $w$, $a$, $b$, etc.)
to stress that these equations relate four-fermion
effective couplings originating from the
{\it same} boson intermediate state;
$n=S$ for scalar mediated decays,
and $n=V$ for vector mediated decays.
These relations hold within any of the three channels,
$(\mu, e)$,
$(\tau, e)$,
and $(\tau, \mu)$.

Moreover,
there are additional equations relating different processes,
such as
\ba\label{eq:cross}
\alpha^n_{\mu_L \tau_L}\ \alpha^n_{e_L \tau_R}
& = &
\alpha^n_{\mu_L \tau_R}\ \alpha^n_{e_L \tau_L}\ ,
\nonumber\\*[3mm]
\alpha^n_{\mu_L \tau_L}\ \alpha^{n\ast}_{e_L \mu_R}
& = &
\alpha^n_{\mu_R \tau_L}\ \alpha^{n\ast}_{e_L \mu_L}\ ,
\\*[3mm]
\alpha^n_{e_L \tau_L}\ \alpha^n_{e_R \mu_L}
& = &
\alpha^n_{e_R \tau_L}\ \alpha_{e_L \mu_L}
\nonumber\ ,
\ea
and
\be
\mbox{\rm Im}\left(
\alpha^n_{e_\epsilon \mu_\lambda}\
\alpha_{e_\epsilon \tau_\gamma}^{n\ast}\
\alpha^n_{\mu_\lambda \tau_\gamma}
\right) = 0
\label{eq:imaggeneral}\ ,
\ee
for any chosen set of chiralities ($\epsilon,\lambda,\gamma$).
Other similar equations may be obtained from these with the help
of Eq.~(\ref{eq:factorization}).
Most of these relations constrain pairs of variables to the
space below a hyperbola.

\subsection{Non-standard $W$ interactions}

In this case we consider only $W$-mediated interactions but
admitting the possibility that the $W$ couples non-universally
to leptons of any chirality.
Then,
\be
g^V_{\epsilon \omega} \equiv
w^V_{\epsilon \omega}\ .
\ee
while all other couplings vanish, leading to $\eta =0$.
The normalization condition $N=1$, implies strong (90\% CL) lower
bounds on the $g^V_{LL}$ couplings:
\be
|g^V_{e_L\mu_L}| > 0.997\ ;
\hspace{3mm}
|g^V_{\mu_L\tau_L}| > 0.83\ ;
\hspace{3mm}
|g^V_{e_L\tau_L}| > 0.87^* \  (0.80) \ ;
\hspace{3mm}
|g^V_{l_L\tau_L}| > 0.90
\label{eq:WA}\ .
\ee
The two $|g^V_{e_L\tau_L}|$ limits correspond to the results obtained using
the $(\xi\delta)_e$ measurement ($\ast$),
or ignoring it (number within brackets).

Since in this case the lower bounds of Eq.~(\ref{eq:WA})
are direct limits on the couplings,
$w^V_{LL}$, of the intermediate boson
under study, we can use the factorization
equation (\ref{eq:factorization}), rewritten in the form,
\be
\left| w^V_{e_R \mu_R} \right|=
\left|
\frac{w^V_{e_L \mu_R}\ w^V_{e_R \mu_L}}{w^V_{e_L \mu_L}}
\right|
< 0.0028
\label{eq:usefulsame}\ ,
\ee
to improve the bound on
$g^V_{e_R \mu_R} = w^V_{e_R \mu_R}$
by an order of magnitude.

For the $(\tau, \mu)$ channel, we can
use the lower bound on $g^V_{\mu_L \tau_L}$,
together with the factorization relations among the couplings
of different channels to get the improved (90\% CL) limits:
\be
\left| w^V_{\mu_R \tau_L} \right| =
\left| \frac{w^V_{\mu_L \tau_L}\ w^{V \ast}_{e_L \mu_R}}
{w^{V \ast}_{e_L \mu_L}} \right|
< 0.060 \ ;
\hspace{20mm}
\left| w^V_{\mu_R \tau_R}\right| =
\left| \frac{w^V_{\mu_L \tau_R}\ w^{V \ast}_{e_L \mu_R}}
{w^{V \ast}_{e_L \mu_L}}\right|
< 0.019 \ .
\label{eq:usefulcrossB}
\ee

Similarly, for the $(\tau, e)$ channel we find
\be
\left| w^V_{e_R \tau_L} \right| =
\left|\frac{w^V_{e_L \tau_L}\ w^V_{e_R \mu_L}}{w^V_{e_L \mu_L}} \right|
< 0.047 \ ;
\hspace{20mm}
\left| w^V_{e_R \tau_R} \right|=
\left| \frac{w^V_{e_L \tau_R}\ w^V_{e_R \mu_L}}{w^V_{e_L \mu_L}}\right|
< 0.013 \ .
\label{eq:usefulcrossC}
\ee
Notice that no information on $(\xi\delta)_e$ has been used here.
Thus, for the case of non-standard $W$-mediated interactions,
the relations among channels developed above allow us to improve
the limits on some couplings by one order of magnitude.

Using the bounds (\ref{eq:usefulcrossB}) and (\ref{eq:usefulcrossC}), the
normalization condition $N=1$ allows us to further improve the (90\% CL)
lower limits on the $w^V_{l_L\tau_L}$ couplings
\be
\left| w^V_{\mu_L\tau_L}\right| > 0.95 \ ; \qquad
\left| w^V_{e_L\tau_L}\right| > 0.96 \ ,
\ee
where the last bound is now independent of the $(\xi\delta)_e$
measurement.

Table~\ref{tab:W_couplings} summarizes the limits on $W$-mediated interactions.

%%%%%%%%%%%%  TABLE  %%%%%%%%%%%%
\begin{table}[hbt]
\centering
\begin{tabular}{||l||l|l|l||}
\hline & \hfil $\mu\to e$\hfil &
\hfil $\tau\to\mu$\hfil &\hfil $\tau\to e$ \hfil
\\\hline\hline
$|w^V_{LL}|$ & $>0.997$ & $>0.95$ & $>0.96$
\\
$|w^V_{RR}|$ & $<0.0028$ & $<0.019$ & $<0.013$
\\
$|w^V_{LR}|$ & $<0.060$ & $<0.31$ & $<0.27$
\\
$|w^V_{RL}|$ & $<0.047$ & $<0.060$ & $<0.047$
\\ \hline
\end{tabular}
\caption{90\% CL limits
for the $w^V_{\epsilon \omega}$ couplings, assuming that any
additional interactions are negligible.}
\label{tab:W_couplings}
\end{table}
%%%%%%%%%%%%%%%%%%%%%%%%%%%%%%%%%

\subsection{SM plus Charged Vector}

If in addition to the SM $W$ boson ($w^V_{LL} \neq 0$,
and all others zero), there is another vector boson with
a mass not too large, then its presence will be constrained by
the effective vector couplings
($a^V_{\epsilon \omega}$) that it generates.
In particular, we have seen in Sect.~\ref{sec:universality}
that differences of
\be
g^V_{LL} = w^V_{LL} + a^V_{LL}
\ee
corresponding to different channels are well constrained
by universality tests.

The general analysis follows the one of the previous case,
except for the fact that
Eqs.~(\ref{eq:usefulsame}), (\ref{eq:usefulcrossB})
and (\ref{eq:usefulcrossC}) do not provide upper bounds on
the single couplings on the left hand side.
Indeed, the lower bound on
$g^V_{e_L \mu_L}$, which affects the sum of the SM with the new
contribution,
does not translate into a lower bound for
$a^V_{e_L \mu_L}$.
This is just a reflection of the fact that the experiments are
consistent with the inexistence of a contribution from a new
vector boson.
Of course, those relations are still useful in the form of
Eqs.~(\ref{eq:cross}), to limit products of couplings.
What we cannot do in this case, is use these relations,
together with the lower bound on $g^V_{e_L \mu_L}$, to place
limits on a single coupling.

For instance,
\ba
|a^V_{e_L \mu_L}\ g^V_{e_R \mu_R}|
= |g^V_{e_L \mu_R}\ g^V_{e_R \mu_L}|
&<& 0.0028 \quad (90\%\;\mbox{\rm CL})\ ,
\nonumber\\
|a^V_{\mu_L \tau_L}\ g^V_{\mu_R \tau_R}|
= |g^V_{\mu_L \tau_R}\ g^V_{\mu_R \tau_L}|
& < &
0.071 \;\;\quad (90\%\;\mbox{\rm CL})\ ,
\\
|a^V_{e_L \tau_L}\ g^V_{e_R \tau_R}| =
|g^V_{e_L \tau_R}\ g^V_{e_R \tau_L}| \,
& < & 0.073 \;\;\quad (90\%\;\mbox{\rm CL})\ . \nonumber
\ea
These equations establish non-trivial constraints since they
involve $a^V_{LL}$, to which we do not have direct
experimental access.
So, in addition to direct bounds on individual magnitudes,
we have also constrained the allowed values to
the space below a hyperbola, in the respective plane.
Of course, there are many such constraints. Here we just want to
illustrate their existence and point out that these constraints
translate into non-trivial information and might be especially
useful in specific models that have a small number of parameters.

\subsection{SM plus Charged Scalar}

In this case $\rho = 3/4$ and
\be
\begin{array}{c}
2\ \eta = \mbox{\rm Re}(w^V_{LL}\ g^{S\ast}_{RR})
\sim \mbox{\rm Re}(g^S_{RR})\ ,
\\
2(1-\xi) = 2 \left(1-\frac{4}{3}\xi\delta\right)
= \left|g^S_{LR}\right|^2 + \left|g^S_{RR}\right|^2 \, .
\end{array}
\ee
The positivity of $(1-\xi)$ leads now to slightly improved
(90\% CL) bounds for the scalar couplings,
\be\label{eq:cs_coup}
\begin{array}{cccc}
\left|g^S_{e_L\mu_R}\right|^2 + \left|g^S_{e_R\mu_R}\right|^2
& =& -0.006\pm 0.016
&< \, 0.023 \ , \\*[2mm]
\left|g^S_{\mu_L\tau_R}\right|^2 + \left|g^S_{\mu_R\tau_R}\right|^2
& =&  -0.46\pm 0.48
&<\,  0.56 \ , \\*[2mm]
\left|g^S_{e_L\tau_R}\right|^2 + \left|g^S_{e_R\tau_R}\right|^2
& =& -0.06\pm 0.50
&<\,  0.79 \ , \\*[2mm]
\left|g^S_{l_L\tau_R}\right|^2 + \left|g^S_{l_R\tau_R}\right|^2
& =& -0.12\pm 0.22
&<\,  0.30 \ .
\end{array}
\ee
The limits on the $(\mu, e)$ couplings are still weaker than the ones in
Table~\ref{tab:mu_couplings}, but the others are stronger than the ones in
Eqs.~(\ref{eq:delta_bounds_CH}) and (\ref{eq:xi_rho}). Only the
bound obtained from $(\xi\delta)_e$ is better.

The information on the low-energy parameter $\eta$ gives the (90\% CL) limits:
\be\label{eq:eta_bounds_s}
\begin{array}{rcl}
-0.057 & < \mbox{\rm Re}\left( w^V_{e_L\mu_L}
g^{S\ast}_{e_R\mu_R}\right)
< & 0.029 \ , \\
-1.03 & < \mbox{\rm Re}\left( w^V_{\mu_L\tau_L}
g^{S\ast}_{\mu_R\tau_R}\right)
< & 0.47 \ .
\end{array}\ee
Assuming lepton universality, Eq.~(\ref{eq:eta_univ})
yields a much better bound on the
$\tau \to l$ couplings:
\be\label{eq:eta_univ_tau}
\left( \begin{array}{c} -0.18 \\ -0.09 \end{array}\right)
 < \mbox{\rm Re}\left( w^V_{l_L\tau_L}
g^{S\ast}_{l_R\tau_R}\right)
< \left( \begin{array}{c} 0.32 \\ 0.12 \end{array}\right)
\qquad\quad \left(\begin{array}{c} \mbox{\rm PDG 94} \\
\mbox{\rm Montreux 94}\end{array}\right) \ ,
\ee
which, however, is still worse than the limit obtained from
$\eta_{\mu\to e}$.

Using the factorization relations, one gets additional limits,
such as
\begin{eqnarray}
|g^S_{e_L \mu_R}\ g^S_{e_R \mu_L}| = |g^S_{e_L \mu_L}\ g^S_{e_R \mu_R}|
&<& 3.6 \times 10^{-2}\ ,
\nonumber\\
|g^S_{e_R \mu_L}\ g^S_{\mu_R \tau_R}|
= |g^S_{e_R \mu_R}\ g^S_{\mu_L \tau_R}|
&<& 0.050 \ .
\end{eqnarray}
improving the limits on the products in the left-hand side of the
equations over the bounds obtainable directly from
Table~\ref{tab:coup_CH}.
At present,
the $\tau_L$ couplings are only constrained by the normalization
condition: $|g^S_{l_\epsilon\tau_L}|<2$ and
$|g^V_{l_L\tau_L}|<1$.

\section{Constraints on new neutral bosons}

In this section we study the possible existence of neutral
bosons violating the leptonic $l$ and $l^\prime$ numbers.
For example, in models with heavier leptons with non-canonical
quantum number assignments, there are non-diagonal $Z^0$
interactions induced by the mixing of the standard leptons with
exotic ones.
In other models, similar couplings with new neutral scalars
arise naturally at levels close to the current experimental
values \cite{barroso}.

Of course, such interactions will also contribute to the well
constrained flavour-violating decays into three charged leptons,
such as $\mu \rightarrow eee$.
The $l^-\to\nu_l l'^- \bar\nu_{l'}$ decays
involve two charged-lepton and two neutrino couplings to the
intermediate boson,
while decays of the type $l^- \to l_1^- l_2^+ l_3^-$
involve four charged-lepton couplings.
Therefore, the two types of decay provide complementary information.
Note, however, that in many models the neutrino and charged-lepton couplings
are related; in such cases, the constraints from the
$l^- \to l_1^- l_2^+ l_3^-$ decays are usually much stronger than those
obtained from the $l^-\to\nu_l l'^- \bar\nu_{l'}$ spectra.

It is easily shown that if, as we are assuming, the final neutrinos
are massless and not observed, one falls back on an effective
hamiltonian like that of Eq.~(\ref{eq:hamiltonian}),
even in the presence of lepton-number nonconservation
\cite{langacker}.
In the appendix,
this is shown explicitly for the case of neutral
boson mediated interactions.
We also include there the derivation
of some formulae useful in this section,
and a discussion of bounds from neutrinoless charged-lepton decays.
In the cases studied here
there are no relations among different channels.

\subsection{SM plus Neutral Vector}

When the decay is mediated by neutral vector bosons,
all the LR and RL couplings vanish and $\rho = 3/4$.
Since there are no tensor couplings, the relevant bounds
on Table~\ref{tab:coup_CH} are also valid in this case.
Moreover, $(1-\xi)$ is now a positive-semidefinite quantity,
which gives the additional (90\% CL) limits,
\be\label{eq:nv_coup}
\begin{array}{ll}
\frac{1}{2} \left|g^S_{e_R\mu_R}\right|^2 + 2
\left|g^V_{e_R\mu_R}\right|^2
&< 0.011 \ , \\*[2mm]
\frac{1}{2} \left|g^S_{\mu_R\tau_R}\right|^2 + 2
\left|g^V_{\mu_R\tau_R}\right|^2
&< 0.28 \ , \\*[2mm]
\frac{1}{2} \left|g^S_{e_R\tau_R}\right|^2 + 2
\left|g^V_{e_R\tau_R}\right|^2
&< 0.39 \ , \\*[2mm]
\frac{1}{2} \left|g^S_{l_R\tau_R}\right|^2 + 2
\left|g^V_{l_R\tau_R}\right|^2
&< 0.15 \ .
\end{array}
\ee
The limits on the $(\mu, e)$ couplings are weaker than the ones in
Table~\ref{tab:mu_couplings},
but the others are stronger than the ones in
Eqs.~(\ref{eq:delta_bounds_CH}) and (\ref{eq:xi_rho}).
Only the bound obtained from $(\xi\delta)_e$ is better.

As usual, we distinguish the SM $W$
and the neutral vector boson contributions to $g^V_{LL}$ by
the letters $w$ and $a$, respectively. Hence,
\be
g^V_{LL} = w^V_{LL} + a^V_{LL}\ .
\ee
As shown in the appendix A.1, the new contributions satisfy the relation
\be
a^S_{LL}\ a^S_{RR} = 4\ a^V_{LL}\ a^V_{RR}\ ,
\label{eq:neutralscalar}
\ee
which yields the 90\% CL bounds
\ba
|a^V_{e_L \mu_L}\ g^V_{e_R \mu_R}|
= \frac{1}{4}\
|g^S_{e_L \mu_L}\ g^S_{e_R \mu_R}|
&<&
9.1 \times 10^{-3}\ ,
\nonumber\\
|a^V_{\mu_L \tau_L}\ g^V_{\mu_R \tau_R}|
= \frac{1}{4}\ |g^S_{\mu_L \tau_L}\ g^S_{\mu_R \tau_R}|
&<& 0.37\ ,
\\
|a^V_{e_L \tau_L}\ g^V_{e_R \tau_R}|
= \frac{1}{4}\ |g^S_{e_L \tau_L}\ g^S_{e_R \tau_R}|  \,
&<& 0.32^* \; (0.44)\ . \nonumber
\ea
Again, these relations yield constraints on $a^V_{LL}$,
to which there is no direct experimental access.

%%%%%%%%%%%%  TABLE  %%%%%%%%%%%%
\begin{table}[hbt]
\centering
\begin{tabular}{||c||c|c|c||}
\hline & \hfil $\mu\to e$\hfil &
\hfil $\tau\to\mu$\hfil &\hfil $\tau\to e$ \hfil
\\\hline\hline
$|\alpha_{l' l}|^2\ \sum_{m,n} |\theta_{mn}|^2$
& $<1.1 \times 10^{-3}$ & $<0.14$ & $<0.20$
\\
$|\beta_{l' l}|^2\ \sum_{m,n} |\theta_{mn}|^2$
& $<7.6 \times 10^{-2}$ &  -----  &  -----
\\ \hline
\end{tabular}
\caption{90\% CL limits
on products of quadratic polynomials in the lepton and neutrino couplings.
If one uses the measured value of $(\xi\delta)_e$,
the number on the last
column will read instead $0.10$.
}
\label{tab:neutvectg}
\end{table}
%%%%%%%%%%%%%%%%%%%%%%%%%%%%%%%%%

Assuming that the neutrinos are not detected,
the neutral-vector-induced effective couplings
may be written as
\ba
a^V_{l'_R l^{\phantom{'}}_R} =
\alpha_{l' l} \left[ \sum_{m,n} |\theta_{mn}|^2 \right]^{1/2}
\hspace{4mm}
& , &
\hspace{4mm}
a^S_{l'_R l^{\phantom{'}}_R} =
-2\ \alpha_{l' l} \left[ \sum_{m,n} |\sigma_{mn}|^2 \right]^{1/2}
\ ,
\nonumber\\
a^V_{l'_L l^{\phantom{'}}_L} =
\beta_{l' l} \left[ \sum_{m,n} |\sigma_{mn}|^2 \right]^{1/2}
\hspace{4mm}
& , &
\hspace{4mm}
a^S_{l'_L l^{\phantom{'}}_L} =
-2\ \beta_{l' l} \left[ \sum_{m,n} |\theta_{mn}|^2 \right]^{1/2}
\ ,
\label{eq:fundneutvect}
\ea
where $\alpha$ ($\beta$) is the hermitian coupling matrix
of the right- (left-) handed charged leptons to the
neutral vector and $\theta$ ($\sigma$) is the
coupling matrix of the right- (left-) handed neutrinos,
in appropriate units (see appendix A.1).
The experimental limits on the effective four-fermion couplings
constrain then these combinations
of the original vector couplings.
We summarize these results in Table~\ref{tab:neutvectg}.
The bounds on the first line remain the same with $\theta$ substituted
by $\sigma$ and
the missing numbers on the second line are due to the lack
of experimental access to $Q_{\tau_L}$.

\subsection{SM plus Neutral Scalars}

Finally, we consider the case in which there is a neutral scalar
contribution to $\mu$ and $\tau$ leptonic decays,
in addition to the SM contribution.
These new contributions vanish for the LL and RR
couplings and satisfy the relations
\be
a^V_{LR} = a^S_{LR} = 2 a^T_{LR}
\hspace{7mm}
;
\hspace{7mm}
a^V_{RL} =  a^S_{RL} = 2 a^T_{RL}\ .
\label{eq:VSneutralscalar}
\ee
This allows us to express everything in terms of the
vector couplings. One gets then the positive definite quantities:
\be
1-{4\over 3}\rho = 1-{4\over 3}\xi\delta =
2 \left( |g^V_{LR}|^2 + |g^V_{RL}|^2 \right) \ ,
\ee
and
\be
\left(1-{4\over 3}\rho\right) + {1\over 2} \left( 1-\xi\right) =
6 |g^V_{RL}|^2 \ .
\ee
The $N=1$ constraint provides the additional relation
\be
1 = |g^V_{LL}|^2 + 2 \left( |g^V_{LR}|^2 + |g^V_{RL}|^2 \right) \ .
\ee
Thus,
$1-{4\over 3}\rho = 1-{4\over 3}\xi\delta = \left( 1 - |g^V_{LL}|^2\right)$,
which gives lower bounds on all $g^V_{LL}$ couplings.
The resulting 90\% CL limits are given in Table~\ref{tab:ns_coup}.

%%%%%%%%%%%%  TABLE  %%%%%%%%%%%%
\begin{table}[hbt]
\centering
\begin{tabular}{||l||l|l|l||l||}
\hline & \hfil $\mu\to e$\hfil &
\hfil $\tau\to\mu$\hfil &\hfil $\tau\to e$ \hfil &
\hfil $\tau\to l$ \hfil
\\\hline\hline
$|g^V_{LL}|$ & $>0.998$ & $>0.95$ & $>0.96$ & $>0.97$
\\
$|g^V_{LR}|$ & $<0.047$ & $<0.22$ & $<0.19$ & $<0.18$
\\
$|g^V_{RL}|$ & $<0.033$ & $<0.16$ & $<0.19$ & $<0.13$
\\ \hline
\end{tabular}
\caption{90\% CL limits
for the $g^n_{\epsilon\omega}$ couplings, taking
$g^n_{RR}=0$, $g^S_{LL}=0$, $g^V_{LR}=g^S_{LR}=2 g^T_{LR}$ and
$g^V_{RL}=g^S_{RL}=2 g^T_{RL}$.}
\label{tab:ns_coup}
\end{table}
%%%%%%%%%%%%%%%%%%%%%%%%%%%%%%%%%

In addition we have the constraints from $\eta$, which
at 90\% CL give
\be\label{eq:eta_bounds_ns}
\begin{array}{rcl}
-0.007 & < \mbox{\rm Re}\left(
g^V_{e_L \mu_R}\ g^{V \ast}_{e_R \mu_L}
\right)
< & 0.004 \ ,
\\
-0.13 & < \mbox{\rm Re}\left(
g^V_{\mu_L \tau_R}\ g^{V \ast}_{\mu_R \tau_L}
\right)
< & 0.06 \ .
\end{array}\ee

These effective couplings may be written in terms of the ones in the
original lagrangian as
\be
a^V_{l'_R l^{\phantom{'}}_L} =
 A_{l' l} \left[ \sum_{m,n} |B_{mn}|^2 \right]^{1/2}
\ ,\qquad
a^V_{l'_L l^{\phantom{'}}_R} =
 A^\ast_{l l'} \left[ \sum_{m,n}
|B_{mn}|^2 \right]^{1/2}\ ,
\ee
where $A$ ($B$) is the coupling matrix of the charged leptons
(neutrinos) to the neutral scalar,
in appropriate units (see appendix A.2).
So, the previous limits contain combined information from the
two sectors.

It is important to emphasize that, within the philosophy
we sustain of discarding intermediate tensor particles
(for they hardly appear in any reasonable model beyond the
SM), this is the only possible source of tensorial terms.
This fact has interesting consequences which we will
explore in the next section.

\section{Opportunities for Physics Beyond the SM}

In Table~\ref{tab:summary}
 we present a summary of the theoretical constraints
imposed on the measured quantities , for the various cases
under study. There, SM denotes that the Standard Model
results are recovered and AS indicates that any sign is allowed.

\begin{table}
\begin{tabular}{|c|c|c|c|c|}
\hline
                                                             &
{\it SM + Charged}                                           &
{\it SM + Charged}	                                     &
{\it SM + Neutral}	                                     &
{\it SM + Neutral}	                                     \\
                                                             &
{\it Vector}                                                 &
{\it Scalar}		                                     &
{\it Vector}		                                     &
{\it Scalar}		                                     \\
                                                             &
{\it Nonstandard W}                                          &
			                                     &
			                                     &
			                                     \\
\hline
$\rho - 3/4$                                                 &
$< 0$                                                        &
SM                                                           &
SM                                                           &
$< 0$                                                        \\
\hline
$\xi - 1$                                                    &
AS                                                           &
$< 0$                                                        &
$< 0$                                                        &
AS                                                           \\
\hline
$(\delta \xi) - 3/4$                                         &
$< 0$                                                        &
$< 0$                                                        &
$< 0$                                                        &
$< 0$                                                        \\
\hline
$\eta$                                                       &
SM                                                           &
AS                                                           &
AS                                                           &
AS                                                           \\
\hline
\end{tabular}
\caption{Theoretical constraints
on the Michel parameters}\label{tab:summary}
\end{table}

It is immediately apparent that $\rho \leq 3/4$ and
$(\delta \xi) < 3/4$ in all cases that we have studied.
Thus one can have new physics and still $\rho$ be equal to
the SM value. In fact, any interaction consisting
of an arbitrary combination of $g^S_{\epsilon \omega}$'s
and $g^V_{\gamma \gamma}$'s yields this result
\cite{FE:90}.
On the other hand, $(\delta \xi)$ will be different from
$3/4$ in any of the cases above providing, in principle,
a better opportunity for the detection of Physics Beyond the SM.

The above features are easy to understand by looking back at
Eqs.~(\ref{eq:michel}) and recalling that the tensor couplings
can only be generated by neutral scalar interactions
(violating individual lepton flavours),
in which case they are proportional to the scalar couplings.
It is easy to see that having two such neutral scalars will
not alter the situation. Indeed, to obtain
$\rho > 3/4$ or $(\delta \xi) > 3/4$ one will also need
the presence of a charged scalar.

Let's then assume that we have a neutral scalar with couplings
\be
a^V_{LR} = a^S_{LR} = 2\ a^T_{LR}
\hspace{7mm}
;
\hspace{7mm}
a^V_{RL} = a^S_{RL} = 2\ a^T_{RL}
\ ,
\ee
and a charged scalar with couplings $b^S_{\epsilon \omega}$.
We obtain,
\ba
\rho - \frac{3}{4}
 & = &
- \frac{3}{4}
\left[ 2 {|a^S_{LR}|}^2 + 2 {|a^S_{RL}|}^2
+ \frac{1}{2}{\rm Re}(a^S_{LR} b^{S \ast}_{LR} + a^S_{RL} b^{S \ast}_{RL})
\right]\ ,
\nonumber\\
({\xi}\delta) - \frac{3}{4}
& = &
- \frac{3}{4}
\left[ \frac{1}{2}{|b^S_{RR}|}^2
+ \frac{1}{8} {|a^S_{LR} - b^S_{LR}|}^2
+ \frac{3}{8} {|a^S_{LR} + b^S_{LR}|}^2
\right.
\nonumber\\
&   &
\quad\left.
+ \frac{3}{2} {|a^S_{LR}|}^2
+ 2 {|a^S_{RL}|}^2
+ \frac{1}{2} {\rm Re}(a^S_{RL} b^{S \ast}_{RL})
\right]\ .
\ea
The first equation shows that $\rho$ might exceed $3/4$
provided that
\be
{\rm Re}
\left( \frac{b^S_{LR}}{a^S_{LR}} \right)
< - 4\ ,
\ee
or
\be
{\rm Re}
\left( \frac{b^S_{RL}}{a^S_{RL}} \right)
< - 4\ .
\ee

As for $(\delta \xi)$, it can only exceed the SM value
through $RL$ couplings, and only if the last equation is
satisfied. Then, detecting $\rho$ greater than the SM value
would mean that there were at least a charged scalar and
a neutral scalar in action. A measurement of $(\delta \xi)$
greater than $3/4$ would then discriminate between $RL$ and
$LR$ couplings.
However, as pointed out before,
a measurement of $(\delta \xi) > 3/4$ must, in general,
be accompanied by a measurement of $\xi > 1$.
If the contrary were to become well established,
we would have detected physics beyond the four-fermion hamiltonian.

\section{Conclusions}
We have used the recent measurements on the Michel parameters
in tau decays to perform a complete, model independent
analysis of the constraints implied for scalar and neutral bosons,
as they exist in most models beyond the SM.
If the new contributions are dominated by
the effect of one such new intermediate boson, relations among the
different couplings arise.

In the case of charged intermediate bosons, these relations involve
couplings from different decays.
If the most important new feature is the coupling of the usual
$W$ boson with right handed leptons, then the data from
muon neutrino scattering off electrons can be used to
improve some of the limits on couplings in tau decays by an
order of magnitude. In the other cases, it constrains products
of couplings of different channels to the space below a
hyperbola. This information will be particularly useful for models
in which these couplings are functions of the same parameters
of the original theory.

In case the dominant new features are provided by the exchange
of flavour violating neutral scalars, there are no relations among
the different channels. The relations within each channel were
derived assuming that the final neutrinos are massless and not
observed. This shows explicitly that the analysis based on the
common four fermion hamiltonian is still valid in this case.
It is shown in the appendix
that, given the current experimental situation,
the bounds obtained from the Michel parameters only compete with
those provided by the decays $l^- \to l_1^- l_2^+ l_3^-$,
in theories were the charged lepton couplings to the intermediate
particle carrying flavour are suppressed by some
(exact or approximate) symmetry.

\section*{Acknowledgements}

This work has been supported in part by CICYT (Spain) under
grant No. AEN-93-0234.
The work of J.P.S. was funded by the E.U. under the
Human Capital and Mobility Program.
He is indebted to G.C. Branco, L. Lavoura and
Instituto Superior T\'ecnico for their
kind hospitality, and to J. Raab and A. Stahl for useful discussions.

\appendix
\section{Flavour-violating neutral-mediated interactions}

\subsection{Neutral Vector Bosons}

We parametrize the interaction of a    %new flavour-violating,
neutral vector boson $V^0_\mu$ with the leptons by
\be
{\cal L} = J^\mu V_\mu^0\ ,
\label{eq:lagrangianneutvect}
\ee
where
\be
J^\mu =
\left( M_V \sqrt{2 \sqrt{2} G_F} \right)
\left[
\bar{l} \gamma^\mu (\alpha \gamma_R + \beta \gamma_L) l
+ \bar{\nu} \gamma^\mu (\theta \gamma_R + \sigma \gamma_L) \nu
\right]\ .
\ee
$M_V$ is the mass of the neutral vector,
$\gamma_{R,L}\equiv (1\pm\gamma_5)/2$ are the chirality projectors,
and $\alpha$, $\beta$, $\theta$ and $\sigma$ are $3 \times 3$
hermitian matrices in the respective flavour spaces.

Since $M_V$ is typically much larger than the energy scale of
$\mu$ and $\tau$ decays, the interaction is effectively that of
four fermions,
\be
{\cal H} = \frac{1}{2 M_V^2} J^\mu J^\dagger_\mu\ ,
\ee
which, after Fierzing, yields an effective interaction
between two neutrinos and two charged leptons of the form,
\ba\label{eq:H_eff_NV}
{\cal H}
& \!\!\! = &\!\!\!
\frac{4 G_F}{\sqrt{2}}
\left[
\alpha_{l' l} \theta_{mn}
(\overline{l'_R} \gamma^\mu \nu^n_{R})
(\overline{\nu^m_R} \gamma_\mu l_R)
+ \beta_{l' l} \sigma_{mn}
(\overline{l'_L} \gamma^\mu \nu^n_{L})
(\overline{\nu^m_L} \gamma_\mu l_L)
\right.
\nonumber\\
& &\quad\;\;
\left.
-2\alpha_{l' l} \sigma_{mn}
(\overline{l'_R} \nu^n_{L})
(\overline{\nu^m_L} l_R)
- 2\beta_{l' l} \theta_{mn}
(\overline{l'_L} \nu^n_{R})
(\overline{\nu^m_R} l_L)
\right]\ .
\ea
As expected, the chirality changing couplings RL and LR
vanish. Notice that the resulting effective couplings satisfy the
relation
\be
a^S_{l'_R l^{\phantom{'}}_R} a^S_{l'_L l^{\phantom{'}}_L} = 4
a^V_{l'_R l^{\phantom{'}}_R} a^V_{l'_L l^{\phantom{'}}_L} \ .
\ee

Since the neutrinos are unobserved,
Eq.~(\ref{eq:H_eff_NV}) reduces to the
hamiltonian (\ref{eq:hamiltonian}),
with effective couplings:
\ba
a^V_{l'_R l^{\phantom{'}}_R} =
\alpha_{l' l}\ \Theta
\hspace{4mm}
& , &
\hspace{4mm}
a^S_{l'_R l^{\phantom{'}}_R} =
-2\ \alpha_{l' l}\ \Sigma
\ ,
\nonumber\\
a^V_{l'_L l^{\phantom{'}}_L} =
\beta_{l' l}\ \Sigma
\hspace{4mm}
& , &
\hspace{4mm}
a^S_{l'_L l^{\phantom{'}}_L} =
-2\ \beta_{l' l}\ \Theta
\ ,
\ea
where
\be
\Theta = \left[ \sum_{m,n} |\theta_{mn}|^2 \right]^{1/2}\ ,
\qquad\quad
\Sigma = \left[ \sum_{m,n} |\sigma_{mn}|^2 \right]^{1/2}\ .
\ee

The lagrangian (\ref{eq:lagrangianneutvect})
also induces flavour-changing decays of $\mu$ and $\tau$ into
three charged leptons.
Neglecting the masses of the final leptons, the corresponding decay widths
can be written as
\be
\Gamma[l^- \to l_1^- l_2^+ l_3^-]
= {m_l^5 G_F^2\over 192\pi^3}\ {\cal C}^l_{l_1 l_2 l_3}\ ,
\ee
where
${\cal C}^l_{l_1 l_2 l_3}$ are quartic polynomials in the
lepton couplings $\alpha$ and $\beta$,
given by
\ba
{\cal C}^l_{l_1 l_2 l_3}
&\!\!\!  = &\!\!\! w_s
\left\{
\left( |\alpha_{l_1 l}|^2 + |\beta_{l_1 l}|^2 \right)
\left( |\alpha_{l_3 l_2}|^2 + |\beta_{l_3 l_2}|^2 \right)
\right. \nonumber\\  &  &\quad
+ \left( |\alpha_{l_3 l}|^2 + |\beta_{l_3 l}|^2 \right)
\left( |\alpha_{l_1 l_2}|^2 + |\beta_{l_1 l_2}|^2 \right)
\\ &  &\quad\left.
 + 2 \mbox{\rm Re}
\left[
\alpha_{l_1 l} \alpha_{l_3 l}^\ast
	\alpha_{l_3 l_2} \alpha_{l_1 l_2}^\ast
+ (\alpha \to \beta)
\right]
\right\} \ ,
\label{eq:appencll1l1l2}
\ea
and $w_S$ denotes the appropriate statistical factor
($w_S=1$ for $l_1\not= l_3$ and $w_S=\frac{1}{2}$ for $l_1= l_3$).
The present experimental bounds on these  % ${\cal C}^l_{l_1 l_2 l_3}$
parameters are given in Table~\ref{tab:neutvectC}.

%%%%%%%%%%%%  TABLE  %%%%%%%%%%%%
\begin{table}[hbt]
\centering
\begin{tabular}{||c||c|c|c|c||}
\hline
$l$ & $l^-_1$ & $l^+_2$ & $l^-_3$ & Bounds on
${\cal C}^l_{l_1 l_2 l_3}$
\\\hline\hline
$\mu$ & $e^-$ & $e^+$ & $e^-$ & $<1.0 \times 10^{-12}$
\\ \hline
   & $e^-$ & $e^+$ & $e^-$ & $<1.9 \times 10^{-5}$
\\
   & $\mu^-$ & $\mu^+$ & $\mu^-$ & $<2.4 \times 10^{-5}$
\\
   & $\mu^-$ & $e^+$ & $\mu^-$ & $<2.0 \times 10^{-5}$
\\
$\tau$   & $e^-$ & $\mu^+$ & $e^-$ & $<1.9 \times 10^{-5}$
\\    %\cline{2-5}
   & $\mu^-$ & $\mu^+$ & $e^-$ & $<2.0 \times 10^{-5}$
\\
   & $e^-$ & $e^+$ & $\mu^-$ & $<1.9 \times 10^{-5}$
\\ \hline
\end{tabular}
\caption{90\% CL limits \protect\cite{PDG:94,CLEO:94}
on products of quartic polynomials in the couplings.}
\label{tab:neutvectC}
\end{table}
%%%%%%%%%%%%%%%%%%%%%%%%%%%%%%%%%

In general, the bounds in Tables~\ref{tab:neutvectg}
and \ref{tab:neutvectC} affect different combinations of the
original couplings. The first gives us combined information on the
charged and neutral lepton sectors, while the second constrains only
the charged lepton sector. However, in many models, these couplings
are related. As an example, let us assume that all diagonal
couplings are of order one with all off-diagonal couplings
suppressed. Then, from Table~\ref{tab:neutvectg} we find
\ba
|\alpha_{e \mu}|^2 & < & 3.7 \times 10^{-4}
\hspace{7mm}
;
\hspace{7mm}
|\beta_{e \mu}|^2 < 2.5 \times 10^{-2}\ ,
\nonumber\\
|\alpha_{\mu \tau}|^2 & < & 4.7 \times 10^{-2}\ ,
\nonumber\\
|\alpha_{e \tau}|^2 & < & 6.7 \times 10^{-2}\ ,
\ea
while the bound on $l \to 3l'$ provides much more
stringent limits,
\ba
|\alpha_{e \mu}|^2 + |\beta_{e \mu}|^2
& < &
3.3 \times 10^{-13}\ ,
\nonumber\\
|\alpha_{\mu \tau}|^2 + |\beta_{\mu \tau}|^2
& < &
8.0 \times 10^{-6}\ ,
\nonumber\\
|\alpha_{e \tau}|^2 + |\beta_{e \tau}|^2 & < &
6.3 \times 10^{-6}\ .
\ea
On the other hand, if we keep the assumption that the diagonal
couplings are dominant,
but there is a hierarchy between the neutrino and lepton
couplings of order
\be
\frac{\theta_{mm}}{\alpha_{ll}} \sim {\cal O}(10^2)\ ,
\ee
then the information in the Michel parameters becomes comparable
with that in $\tau \to 3e$ and $\tau \to 3\mu$ decays.
Of course, this last case will only seem `natural' in the presence
of some (maybe approximate) symmetry suppressing the charged-lepton
couplings to the new neutral vector.

\subsection{Neutral Scalar Bosons}

The interaction of a new flavour-violating,
neutral scalar boson $S^0$ with the leptons can be written as
\be
{\cal L} = J\ S^0\ ,
\label{eq:lagrangianneutscal}
\ee
where
\be
J =
\left( M_S \sqrt{\sqrt{2} G_F} \right)
\left[
\bar{l}  (A^\dagger \gamma_R + A \gamma_L) l
+ \bar{\nu} (B^\dagger \gamma_R + B \gamma_L) \nu
\right]\ ,
\ee
$M_S$ is the mass of the neutral scalar,
and $A$ and $B$ are $3 \times 3$
matrices in the respective flavour spaces.
Again, since $M_S$ is typically much larger than the energy scale of
$\mu$ and $\tau$ decays, the interaction is effectively of
the four fermion type.
After Fierzing, the $\bar l^i l^j \bar\nu^m\nu^n$ interaction is described
by the hamiltonian (\ref{eq:hamiltonian}),
with effective couplings:
\ba
g^S_{l^i_R l^j_L} = 2 g^T_{l^i_R l^j_L} = A_{ij} B_{mn} \ , \quad
&&
g^V_{l^i_R l^j_L} = A_{ij} B_{nm}^\ast \ ,
\nonumber\\
g^S_{l^i_L l^j_R} = 2 g^T_{l^i_L l^j_R} = A_{ji}^\ast B_{nm}^\ast \ ,
&&
g^V_{l^i_L l^j_R} = A_{ji}^\ast B_{mn} \ .
\ea
The LL and RR couplings vanish.
Notice the relation
\be
g^S_{l^i_R l^j_L} g^S_{l^i_L l^j_R} = g^V_{l^i_R l^j_L} g^V_{l^i_L l^j_R} \ .
\ee

Since the neutrinos are unobserved, the
measurable vector and scalar effective couplings are also related:
\ba
g^V_{l'_R l^{\phantom{'}}_L} = g^S_{l'_R l^{\phantom{'}}_L} =
2 g^T_{l'_R l^{\phantom{'}}_L} =
A_{l' l}\ \Omega \ ,
\nonumber\\
g^V_{l'_L l^{\phantom{'}}_R} = g^S_{l'_L l^{\phantom{'}}_R} =
2 g^T_{l'_L l^{\phantom{'}}_R} =
A^\ast_{l l'}\ \Omega\ ,
\ea
where
\be
\Omega=\left[ \sum_{m,n} |B_{mn}|^2 \right]^{1/2}\ .
\ee

This new flavour-changing neutral scalar also contributes
to the decays $l^- \to l_1^- l_2^+ l_3^-$.
The corresponding decay widths are given by
\be
\Gamma[l^- \to l_1^- l_2^+ l_3^-]
= {m_l^5 G_F^2\over 192\pi^3} \ {\cal D}^l_{l_1 l_2 l_3}\ ,
\ee
where the ${\cal D}^l_{l_1 l_2 l_3}$ are quartic polynomials in the
charged-lepton couplings:
\ba
{\cal D}^l_{l_1 l_2 l_3}
&\!\!\! = &\!\!\! w_S
\left\{
\left( |A_{l_1 l}|^2 + |A_{l l_1}|^2 \right)\
\left( |A_{l_2 l_3}|^2 + |A_{l_3 l_2}|^2 \right)
\right. \nonumber\\ &   &\quad \left.
+ \left( |A_{l_3 l}|^2 + |A_{l l_3}|^2 \right)\
\left( |A_{l_1 l_2}|^2 + |A_{l_2 l_1}|^2 \right)
\right. \\ &   &\quad \left.
- \mbox{\rm Re}\left[
A_{l_1 l} A_{l_3 l}^\ast A_{l_3 l_2} A_{l_1 l_2}^\ast
+ A_{l l_1}^\ast A_{l l_3} A_{l_2 l_3}^\ast A_{l_2 l_1}
\right] \right\}
\label{eq:appenldll1l1l2}\ . \nonumber
\ea
For the case of $\mu \to 3e$, this combination of couplings
is greatly simplified,
\be
{\cal D}^\mu_{e e e}= {3\over 2}
\left( |A_{\mu e}|^2 + |A_{e \mu}|^2 \right)\
|A_{e e}|^2 \ .
\ee
Since we have taken identical normalizations, the experimental bounds
on these parameters are the same as those in
Table~\ref{tab:neutvectC}.

Again,
unless in the specific model one is studying
some of the couplings to the charged leptons
are suppressed with respect to the neutrino couplings, the
bounds from the decays $l^- \to l_1^- l_2^+ l_3^-$ will be
much stronger than those from $l^- \to l^{'-} \nu \bar{\nu}$.

% \newpage

\vspace{5mm}
% \newpage

\end{document}